\documentstyle[amssymb,preprint,aps,prl,epsf]{revtex}
\newcommand{\mett}{\mbox{$E\!\!\!\!/_{T}$}}
\newcommand{\met}{\mett}
\def \Et {E_T}
\def \Etg {E_T^{\gamma}}

\def\Gravitino{\widetilde{G}}
\newcommand{\eeggmett}{ee\gamma\gamma\mett}
\def\pbarp{p{\bar p}}

\newcommand{\etal}{{\em et al.}}

\def\ppbar{p{\bar p}}
\newcommand{\invpb}{\rm pb^{-1}} 
\newcommand{\mysect}[1]{}

\newcommand{\ptt}{\mbox{$p_T$}}
\newcommand{\NONE}{\mbox{$\widetilde{\chi}_1^0$}}
\newcommand{\NTWO}{\mbox{$\widetilde{\chi}_2^0$}}
\newcommand{\CONE}{\mbox{$\widetilde{\chi}_1^{\pm}$}}
\newcommand{\CONEP}{\mbox{$\widetilde{\chi}_1^{+}$}}
\newcommand{\CONEM}{\mbox{$\widetilde{\chi}_1^{-}$}}
\newcommand{\none}{\NONE}
\newcommand{\ntwo}{\NTWO}
\newcommand{\cone}{\CONE}

\def\Journal#1#2#3#4{{#1} {\bf #2}, #3 (#4)}

\def\PRL{\rm Phys. Rev. Lett.}
\def\PRD{{\rm Phys. Rev.} D}

\begin{document} 

%
%
\begin{center}
{\bf \Large
Search for Anomalous Production of Diphoton Events with
Missing Transverse Energy at CDF
and Limits on Gauge--Mediated Supersymmetry--Breaking Models
}

\end{center}
%
%

\renewcommand{\baselinestretch}{1.0}
\normalsize
\font\eightit=cmti8
\def\r#1{\ignorespaces $^{#1}$}
\hfilneg
\begin{sloppypar}
\noindent
D.~Acosta,\r {16} J.~Adelman,\r {12} T.~Affolder,\r 9 T.~Akimoto,\r {54}
M.G.~Albrow,\r {15} D.~Ambrose,\r {43} S.~Amerio,\r {42}  
D.~Amidei,\r {33} A.~Anastassov,\r {50} K.~Anikeev,\r {31}
% ...
\end{sloppypar}
\vskip .026in
\begin{center}
(CDF Collaboration)
\end{center}

\vskip .026in
\begin{center}
\r 1  {\eightit Institute of Physics, Academia Sinica, Taipei, Taiwan 11529, 
Republic of China} \\
\r 2  {\eightit Argonne National Laboratory, Argonne, Illinois 60439} \\
\r 3  {\eightit Institut de Fisica d'Altes Energies, Universitat Autonoma
de Barcelona, E-08193, Bellaterra (Barcelona), Spain} \\
\r 4  {\eightit Istituto Nazionale di Fisica Nucleare, University of Bologna,
I-40127 Bologna, Italy} \\
\r 5  {\eightit Brandeis University, Waltham, Massachusetts 02254} \\
\r 6  {\eightit University of California at Davis, Davis, California  95616} \\
\r 7  {\eightit University of California at Los Angeles, Los 
Angeles, California  90024} \\
\r 8  {\eightit University of California at San Diego, La Jolla, California  92093} \\ 
\r 9  {\eightit University of California at Santa Barbara, Santa Barbara, California 
93106} \\ 
\r {10} {\eightit Instituto de Fisica de Cantabria, CSIC-University of Cantabria, 
39005 Santander, Spain} \\
\r {11} {\eightit Carnegie Mellon University, Pittsburgh, PA  15213} \\
\r {12} {\eightit Enrico Fermi Institute, University of Chicago, Chicago, 
Illinois 60637} \\
\r {13}  {\eightit Joint Institute for Nuclear Research, RU-141980 Dubna, Russia}
\\
\r {14} {\eightit Duke University, Durham, North Carolina  27708} \\
\r {15} {\eightit Fermi National Accelerator Laboratory, Batavia, Illinois 
60510} \\
\r {16} {\eightit University of Florida, Gainesville, Florida  32611} \\
\r {17} {\eightit Laboratori Nazionali di Frascati, Istituto Nazionale di Fisica
               Nucleare, I-00044 Frascati, Italy} \\
\r {18} {\eightit University of Geneva, CH-1211 Geneva 4, Switzerland} \\
\r {19} {\eightit Glasgow University, Glasgow G12 8QQ, United Kingdom}\\
\r {20} {\eightit Harvard University, Cambridge, Massachusetts 02138} \\
\r {21} {\eightit The Helsinki Group: Helsinki Institute of Physics; and Division of
High Energy Physics, Department of Physical Sciences, University of Helsinki, FIN-00044, Helsinki, Finland}\\
\r {22} {\eightit Hiroshima University, Higashi-Hiroshima 724, Japan} \\
\r {23} {\eightit University of Illinois, Urbana, Illinois 61801} \\
\r {24} {\eightit The Johns Hopkins University, Baltimore, Maryland 21218} \\
\r {25} {\eightit Institut f\"{u}r Experimentelle Kernphysik, 
Universit\"{a}t Karlsruhe, 76128 Karlsruhe, Germany} \\
\r {26} {\eightit High Energy Accelerator Research Organization (KEK), Tsukuba, 
Ibaraki 305, Japan} \\
\r {27} {\eightit Center for High Energy Physics: Kyungpook National
University, Taegu 702-701; Seoul National University, Seoul 151-742; and
SungKyunKwan University, Suwon 440-746; Korea} \\
\r {28} {\eightit Ernest Orlando Lawrence Berkeley National Laboratory, 
Berkeley, California 94720} \\
\r {29} {\eightit University of Liverpool, Liverpool L69 7ZE, United Kingdom} \\
\r {30} {\eightit University College London, London WC1E 6BT, United Kingdom} \\
\r {31} {\eightit Massachusetts Institute of Technology, Cambridge,
Massachusetts  02139} \\   
\r {32} {\eightit Institute of Particle Physics: McGill University,
Montr\'{e}al, Canada H3A~2T8; and University of Toronto, Toronto, Canada
M5S~1A7} \\
\r {33} {\eightit University of Michigan, Ann Arbor, Michigan 48109} \\
\r {34} {\eightit Michigan State University, East Lansing, Michigan  48824} \\
\r {35} {\eightit Institution for Theoretical and Experimental Physics, ITEP,
Moscow 117259, Russia} \\
\r {36} {\eightit University of New Mexico, Albuquerque, New Mexico 87131} \\
\r {37} {\eightit Northwestern University, Evanston, Illinois  60208} \\
\r {38} {\eightit The Ohio State University, Columbus, Ohio  43210} \\  
\r {39} {\eightit Okayama University, Okayama 700-8530, Japan}\\  
\r {40} {\eightit Osaka City University, Osaka 588, Japan} \\
\r {41} {\eightit University of Oxford, Oxford OX1 3RH, United Kingdom} \\
\r {42} {\eightit University of Padova, Istituto Nazionale di Fisica 
          Nucleare, Sezione di Padova-Trento, I-35131 Padova, Italy} \\
\r {43} {\eightit University of Pennsylvania, Philadelphia, 
        Pennsylvania 19104} \\   
\r {44} {\eightit Istituto Nazionale di Fisica Nucleare, University and Scuola
               Normale Superiore of Pisa, I-56100 Pisa, Italy} \\
\r {45} {\eightit University of Pittsburgh, Pittsburgh, Pennsylvania 15260} \\
\r {46} {\eightit Purdue University, West Lafayette, Indiana 47907} \\
\r {47} {\eightit University of Rochester, Rochester, New York 14627} \\
\r {48} {\eightit The Rockefeller University, New York, New York 10021} \\
\r {49} {\eightit Istituto Nazionale di Fisica Nucleare, Sezione di Roma 1,
University di Roma ``La Sapienza," I-00185 Roma, Italy}\\
\r {50} {\eightit Rutgers University, Piscataway, New Jersey 08855} \\
\r {51} {\eightit Texas A\&M University, College Station, Texas 77843} \\
\r {52} {\eightit Texas Tech University, Lubbock, Texas 79409} \\
\r {53} {\eightit Istituto Nazionale di Fisica Nucleare, University of Trieste/\
Udine, Italy} \\
\r {54} {\eightit University of Tsukuba, Tsukuba, Ibaraki 305, Japan} \\
\r {55} {\eightit Tufts University, Medford, Massachusetts 02155} \\
\r {56} {\eightit Waseda University, Tokyo 169, Japan} \\
\r {57} {\eightit Wayne State University, Detroit, Michigan  48201} \\
\r {58} {\eightit University of Wisconsin, Madison, Wisconsin 53706} \\
\r {59} {\eightit Yale University, New Haven, Connecticut 06520} \\
\end{center}
 
\renewcommand{\baselinestretch}{1.5}
\normalsize

\clearpage
\begin{abstract}

We present the results of a search for anomalous production of diphoton events 
with large missing transverse energy using the Collider Detector at Fermilab. In 
$202$~$\invpb$ of $\ppbar$ collisions at $\sqrt{s}=1.96$~TeV we observe no 
candidate events, with an expected standard model background of 
\mbox{$0.27\pm0.07({\rm stat})\pm0.10({\rm syst})$} events. The results exclude 
a lightest chargino of mass less than 167~GeV/$c^2$, and lightest neutralino of mass less than  
93~GeV/$c^2$ at 95\% C.L. in a gauge--mediated supersymmetry--breaking model 
with a light gravitino.


\end{abstract}
\begin{center}
\vspace*{0.2in}
\hspace*{1.0in} PACS numbers 13.85Rm, 13.85Qk, 14.80.-j,14.80.Ly
\vspace*{0.2in}
\end{center}

\clearpage
%
%
\newcounter{myctr}
\mysect{}

The standard model (SM)~\cite{SMRef} of elementary particles has been enormously
successful, but it is incomplete. For theoretical reasons~\cite{OurGMSB,SUSY 
Workshop}, and because of the `$ee\gamma\gamma$+missing transverse energy 
(\mett)'~\cite{defs} candidate event recorded by the CDF detector in Run~I~\cite{ggmet}, 
there is a compelling rationale to search  in high--energy collisions for the production of heavy 
new particles that decay producing the signature of $\gamma\gamma+\mett$.
Of particular theoretical interest are supersymmetric (SUSY) models with gauge--mediated SUSY--breaking
(GMSB). Characteristically, the effective SUSY--breaking scale ($\Lambda$) can be
as low as 100 TeV, the
lightest SUSY particle is a light
gravitino ($\Gravitino$) that  is assumed to be stable, and the SUSY particles have masses that may make them 
accessible at Tevatron energies~\cite{OurGMSB}. In these models 
the visible signatures are determined by the properties of the
next--to--lightest SUSY particle (NLSP) that may be, for example, a
slepton or the lightest neutralino ($\none$).
In the GMSB model investigated here, the NLSP is a $\none$ decaying
almost exclusively to a photon and a $\Gravitino$ that
penetrates the detector without interacting, producing \mett. 
SUSY particle production at the Tevatron is predicted to be dominated by pairs of the
lightest chargino ($\cone$) and by associated production of a $\cone$ and
the next--to--lightest neutralino ($\ntwo$). Each gaugino pair cascades
down to two $\NONE$'s, leading to a final state of
$\gamma\gamma+\mett+X$, where $X$ represents any other final state
particles.

%
%
\mysect{}

In this paper we summarize~\cite{Minsuk} a search for
anomalous production of inclusive $\gamma\gamma+\mett+X$ events in
data corresponding to an integrated luminosity of 202~$\pm$~12~$\invpb$~\cite{lumi} of $\ppbar$ collisions at
$\sqrt{s}=1.96$~TeV using the CDF~II detector~\cite{detector}. We
examine events with two isolated photons with \mbox{$|\eta|\lesssim
1.0$} and \mbox{$\Etg>13$~GeV} for the presence of large 
$\mett$.  This work extends a previous CDF search~\cite{ggmet} for SUSY in this channel
by using an upgraded detector, a higher $\ppbar$ center-of-mass energy, and a larger data sample. The analysis
selection criteria have been re-optimized to  maximize, {\it a priori}, the
expected sensitivity to GMSB SUSY based only on the background expectations and the predictions 
of the model.
Similar searches for diphoton + \mett\ events have been performed elsewhere~\cite{OtherResults}.


%
%
\mysect{}

We briefly describe the aspects of the CDF~II detector relevant to this 
analysis. The magnetic spectrometer consists of  tracking devices inside 
the \mbox{3-m} diameter, 5-m long superconducting solenoid magnet operating at 
1.4~T. A 90-cm long silicon micro-strip vertex detector, consisting of one single--sided layer and six double--sided layers, with an additional double--sided layer 
at large $\eta$, surrounds the beam pipe. Outside the silicon detector, a 3.1-m 
long drift chamber with 96 layers of sense wires is used with the silicon 
detector to determine the momenta of charged particles and the $z$ position of 
the $\pbarp$ interaction ($z_{\rm vertex}$). The calorimeter, 
constructed of projective towers, each with an electromagnetic and hadronic 
compartment, is divided into a central barrel that surrounds the solenoid coil 
(\mbox{$|\eta|<1.1$}) and a pair of `end-plugs' that cover the region 
\mbox{$1.1<|\eta|<3.6$}. The hadronic compartments of the calorimeter 
are also used to provide a measurement of the arrival time of the particles 
depositing energy in each tower. Wire chambers with cathode--strip readout (the 
CES system), located at shower maximum in the central electromagnetic 
calorimeter, give 2-dimensional profiles of showers.  A system of proportional 
wire chambers in front of the central electromagnetic calorimeters (the CPR 
system) uses the one-radiation-length-thick magnet coil as a `preradiator' to 
determine whether showers start before the calorimeter~\cite{photons}. Muons are 
identified with a system of planar drift chambers situated
 outside the calorimeters in 
the region \mbox{$|\eta|<1.0$}.

\newcommand{\oldsvx}{
 A 90-cm long silicon micro-strip vertex detector (SVX), with one single--sided layer and six double--sided layers in the central region, and an 
additional layer with extended coverage to large $\eta$, surrounds the beam 
pipe. Outside the SVX, a 3.1-m long drift chamber with 96 layers of sense wires 
is used with the SVX to determine the momenta of charged particles and the $z$ 
position of the $\pbarp$ interaction ($z_{\rm vertex}$).
}


%
%
\mysect{}

We select candidate events using both online (during data
taking) and offline selection requirements. Online, events are
selected for the presence of two photon candidates, identified by the
three-level trigger as two isolated electromagnetic clusters~\cite{photons}
with $\Etg> 12$~GeV, or two electromagnetic clusters
with $\Etg>18$~GeV and no isolation requirement. 
The offline event selection requirements for the diphoton candidate sample
are designed to reduce electron and jet/$\pi^0$ backgrounds while accepting well-measured diphoton candidates. 
We require two 
central (approximately $0.05<|\eta|<1.0$) 
electromagnetic clusters that: 
a)~have \mbox{$\Etg>13$~GeV}; 
b)~are not near the boundary in $\phi$ of a calorimeter tower~\cite{Fiducial}; 
c)~have the ratio of hadronic to electromagnetic energy, 
Had/EM, $< 0.055+0.00045\cdot E^{\gamma}$(GeV$^{-1}$); 
d)~have no tracks, or only one track with \mbox{$\ptt <1$~GeV/$c$},
extrapolating to the towers of the cluster; 
e)~are isolated in the calorimeter and tracking chamber~\cite{iso};
f)~have a shower shape in the CES consistent with a single photon;
g)~have no other significant energy deposited nearby in the CES.

%
%
\mysect{}

To minimize the number of events with  large \mett\ due to calorimeter energy 
mis-measurement, we correct for jet ($j$) energy loss in cracks between detector 
components and for nonlinear calorimeter response~\cite{jets}. To avoid any 
remaining cases where a jet is not fully measured by the calorimeter, we remove 
events based on the azimuthal opening angle between the \mett\ direction
and the $\phi$ of any jet with uncorrected \mbox{$\Et>10$~GeV}, $\Delta\phi(\mett,j)$. We require all events
to have $10^\circ < \Delta\phi(\mett,j) < 170^\circ$. To reduce 
beam--related and cosmic--ray backgrounds we require a good vertex with $|{z_{\rm vertex}}|<60$~cm and 
reject events with significant energy out-of-time with the 
collision~\cite{ETOUT}. These backgrounds can also produce \mett\ 
equal in magnitude and opposite in direction to a 
photon, or to the vector sum of the momenta of two photons if they are nearby in 
$\phi$. In this case an event is rejected if there are potential cosmic--ray 
hits in the muon chamber, within 30 degrees of the photon, that are not matched 
to any track. Events are also rejected if there is a pattern of energy in the 
calorimeter indicative of beam--related backgrounds~\cite{beam-halo}. A sample 
of 3,306 diphoton events pass all candidate selection requirements. The 
\mett\ requirement, \mett~$>$~45~GeV, is determined by the final optimization 
procedure that is discussed below, after a more complete description of the backgrounds. 


%
%
\mysect{}

Before the \mett\ requirement, the diphoton 
candidate sample is dominated by QCD interactions producing combinations of photons and jets 
faking photons. In each case only small measured \mett\ is expected, due mostly 
to energy measurement resolution effects.  Standard CDF 
techniques~\cite{photons} are used to estimate the individual contributions for 
the sample to be $47 \pm 6\%$ $\gamma j$, $29\pm 4\%$ $\gamma\gamma$, and 
$24\pm4\%$ $jj$ production.  To estimate the shape of the \mett\ distribution of 
this background we use a control sample of similarly-produced events that have the 
same calorimetric response and resolution. We select 7,806 events that pass 
the same photon $E_T$, $z_{\rm vertex}$, fiducial, $\Delta\phi(\mett$,$j)$, 
beam--related and cosmic--ray background selection requirements, but are allowed 
to satisfy looser photon identification and isolation 
requirements~\cite{LooserCuts}. If an event is in the diphoton candidate sample 
it is rejected from the control sample. The contribution from $e\gamma$ events, 
discussed below, is also subtracted from the control sample. Since the \mett\ 
resolution for a given event is a function of the sum of all the transverse 
energy in the event ($\Sigma{\Et}$), and we observe a small difference between the
$\Sigma{\Et}$ distributions of the diphoton candidate and control samples, we 
correct the \mett\ in the control sample for this difference~\cite{MetMethod}. 
To predict the number of events with large \mett, we normalize the corrected 
control sample distribution to the number of diphoton candidate events in the 
region $\mett < 20$ GeV, and fit the spectrum above 10~GeV to a double 
exponential. We predict $0.01\pm 0.01 ({\rm stat}) \pm 0.01 ({\rm syst})$ events 
with \mett$~>~45$~GeV, where the uncertainty is dominated by differences in the 
predictions using various control sample selection requirements, the choice of 
fit function, and the statistical uncertainties of the sample.


%
%
\mysect{}

Events with an electron and a photon candidate ($W\gamma\rightarrow e\nu\gamma$, 
$Wj\rightarrow e\nu\gamma_{fake}$, $Z\gamma\rightarrow ee\gamma$, etc.) can 
contribute to the diphoton candidate sample when the electron track is lost (by 
tracking inefficiency or bremsstrahlung) to create a fake photon. For $W$ decays 
large \mett\ can come from the neutrinos. This background is estimated using 
$e\gamma$ events from the data.  The diphoton triggers accept electromagnetic 
clusters with tracks so they provide an efficient and unbiased sample of these 
events. We find 462 $e\gamma$ events before the \mett\ requirement. Examining a 
$Z\rightarrow ee$ sample, we estimate $1.0\pm 0.4$\% of electrons will pass the 
diphoton candidate sample requirements, including charged track rejection. By 
multiplying the number of observed $e\gamma\mett$ events by the probability that 
an electron fakes a photon, we estimate $0.14 \pm0.06 ({\rm stat}) \pm 0.05 
({\rm syst})$ background events in the sample with \mett~$>~$45~GeV. The 
uncertainty is dominated by the statistical uncertainty in the fake rate and the 
uncertainty in the purity of the $e\gamma$ sample.


Beam--related sources and cosmic rays overlapped with a SM event can contribute 
to the background by producing spurious energy deposits that in turn affects 
the measured \mett. While the rate at which these events contribute to the 
diphoton candidate sample is low, most contain large \mett. The spurious 
clusters can pass photon cuts. The dominant contribution actually comes from 
sources that produce two photon candidates at once, such as a cosmic muon 
undergoing bremsstrahlung twice. This background is estimated from the data 
using a sample of events with no primary collision and two electromagnetic 
clusters, multiplied by the rate that clusters from cosmic rays pass the 
diphoton candidate sample requirements. Backgrounds where only one of the 
photons, or only the \mett, is from a non-collision source, are estimated to be 
negligible. The total number of events expected from non--collision sources in 
the $\mett~>~45$~GeV sample is $0.12\pm 0.03 ({\rm stat}) \pm 0.09 ({\rm 
syst})$. The uncertainty includes the uncertainty in the rate that spurious 
clusters pass the diphoton selection requirements and takes into account the 
statistics and purity of the sample of events with no primary collision.

%
%
\mysect{}

The \mett\ distribution of the diphoton candidate sample, see Figure~\ref{Data 
Plot}, shows good agreement with that from the expected backgrounds. 
Table~\ref{found} summarizes the number of observed events and predicted 
backgrounds with four different \mett\ requirements. There are no events with 
$\mett >45$~GeV.

%
%

%
%

%
\mysect{}

Since there is no evidence for events with 
anomalous \mett\ in the diphoton
candidate sample, we set limits on new particle production from GMSB using the
parameters suggested in Ref.~\cite{snowmass}. 
To estimate the acceptance for this scenario we generate 
GMSB events using 
ISAJET~\cite{isajet} with CTEQ5L parton distribution
functions~\cite{cteq5l}.
The production cross sections from ISAJET are corrected by a 
$K$-factor of approximately 1.2 to match the next-to-leading order 
(NLO) prediction~\cite{kfactor}. 
We process the events through the GEANT-based~\cite{GEANT} 
detector simulation, and correct the resulting efficiency 
with information from data measurements.

\mysect{}

Since electrons and photons interact similarly in the calorimeter we investigate 
the efficiency of the photon identification and isolation selection criteria by 
using a control sample of electrons from $Z\rightarrow ee$ events. Separate 
efficiency estimates comparing data and detector simulation agree to within 3\%. 
Using the simulation we estimate that if a photon within the fiducial portion 
of the detector is isolated, it has an 80\% probability of passing the 
identification and isolation criteria.  However, the isolation energy of the 
photons is predicted from the Monte Carlo to be a strong function of the SUSY 
scale due to the number and energy of the extra jets produced. We find, for 
example, the single--photon efficiency to be reduced to 62\% at 
M$_{\cone}$=170~GeV$/c^2$. This has a significant impact on the sensitivity. We 
find that the fraction of generated signal events passing all the selection 
requirements, including \mett~$>$~45~GeV, rises linearly from 3.5\% at 
M$_{\cone}$=100~GeV$/c^2$ to approximately 8\% at 180~GeV$/c^2$. It remains 
roughly flat for larger masses due to the increasing inefficiency of the 
$\Delta\phi(\mett$,$j)$ selection requirement. The relative systematic 
uncertainty in the efficiency of the photon identification and isolation 
requirements is approximately 6.5\% per photon. Other significant uncertainties in 
the Monte Carlo model predictions are from initial/final state radiation (10\%), 
$Q^2$ of the interaction (3\%) and uncertainty in parton distribution functions 
(5\%). Combining these numbers with the 6\% luminosity uncertainty gives a total 
relative systematic uncertainty of 18\%.


\newcommand{\origzstuff}{

s offsets the additional acceptance due
to kinematic factors. 

To further investigate the efficiency, by ignoring the electron's track in $Z\rightarrow ee$ events,
we can create a control sample of 
electromagnetic showers~\cite{photons}.  We estimate 
that an isolated photon hitting the fiducial portion of the detector
has a 77\% probability of passing the identification and isolation
criteria.  However, since the isolation energy 
of the photons is
predicted from the Monte Carlo to be a strong
}


%
%
\mysect{}

The kinematic selection requirements defining the final data sample are 
determined by a study to optimize the expected limit, $i.e.$, without looking at 
the signal region data. To compute the expected 95\% confidence level (C.L.) 
cross section upper limit we combine the predicted signal and background 
estimates with the systematic uncertainties using a Bayesian 
method~\cite{Conway} and follow the prescription described in Ref.~\cite{Boos}. 
The expected limits are computed as a function of \mett, photon $\Et$, and 
$\Delta\phi(\mett$,$j)$ selection requirements. We find that the best limit is 
predicted with the selection described above for the diphoton candidate sample, 
and \mett$>$~45~GeV.  The statistical analysis indicates that the most probable 
expected result, in the absence of a signal, would be an exclusion of M$_{\CONE}$ less than 161~GeV/$c^2$ and 
M$_{\none}$ less than 86~GeV/$c^2$.  


%
%
\mysect{}

\begin{figure}
\begin{centering}
\epsfysize=6.0in
\epsfbox{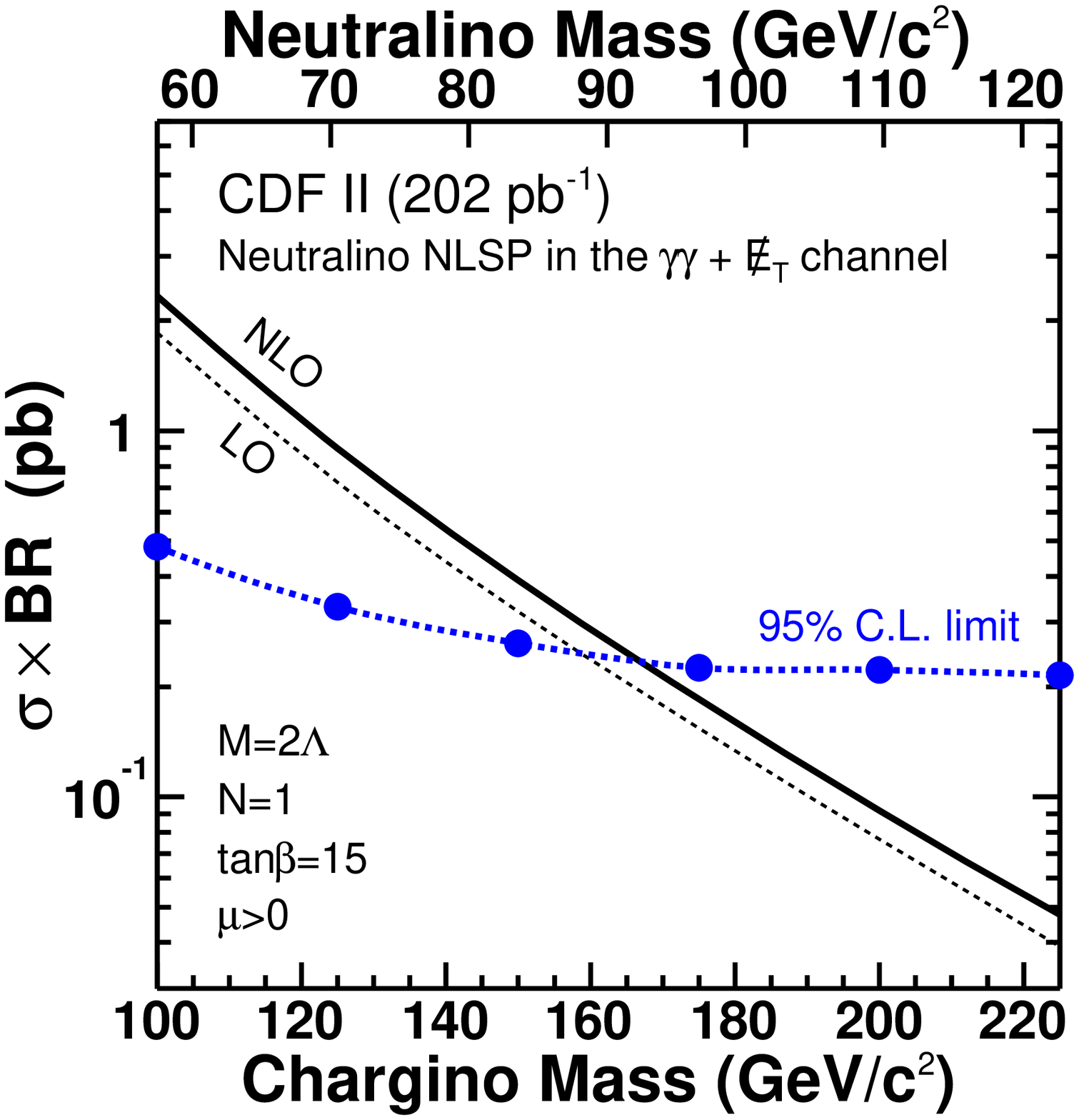}

\caption{The 95\% C.L. upper limits on the total production cross section times 
branching ratio versus M$_{\CONE}$ and M$_{\none}$ for the light gravitino 
scenario using the parameters proposed in~\protect\cite{snowmass}.  The lines 
show the experimental limit and the LO and NLO theoretically predicted cross 
sections. We set limits of M$_{\cone}>167$~GeV/$c^2$ and \mbox{M$_{\none}>93$~GeV/$c^2$} 
at 95\% C.L. }

\label{Limit Plot}
\end{centering}
\end{figure}

In the data signal region, with \mett$>$~45~GeV, we observe zero events. Taking 
into account the 18\% systematic uncertainty we set a 95\% C.L. upper limit of 
3.3 signal events. Figure~\ref{Limit Plot} shows the observed cross section 
limits as a function of M$_{\CONE}$ and M$_{\none}$ along with the theoretical LO and NLO 
production cross sections.  Using the NLO predictions we set a limit of 
M$_{\cone} > 167$~GeV/$c^2$ at 95\% C.L. From mass relations in the model, we 
equivalently exclude \mbox{M$_{\none}<93$~GeV/$c^2$} and $\Lambda <$ 69~TeV.

%
%

\begin{figure}
\begin{centering}
\epsfysize=6.0in
\vspace*{-0.4cm}
\epsfbox{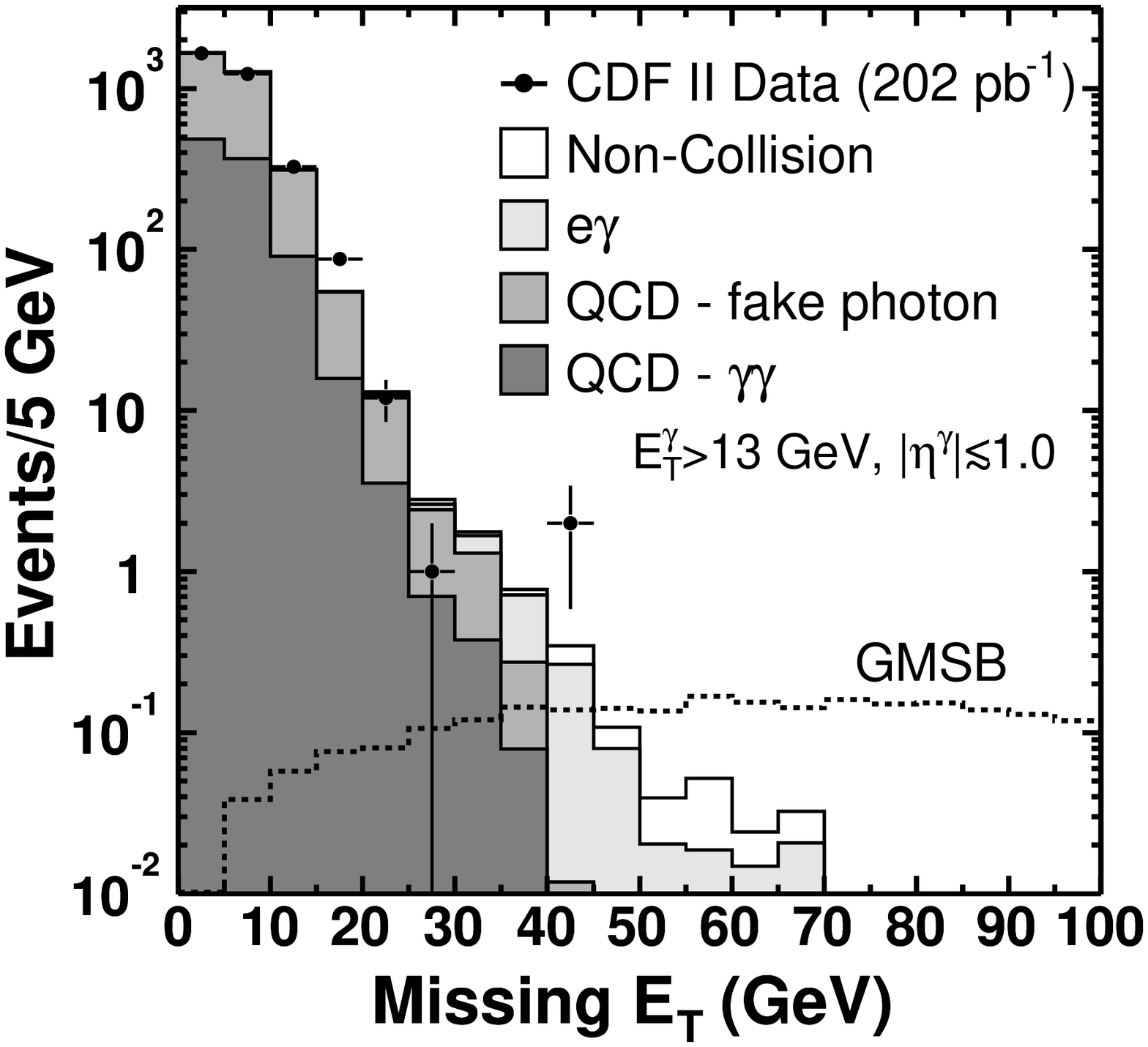}
\caption{
The \mett\ spectrum for events with two isolated central
photons with $\Etg>13$~GeV 
and \protect{\mbox{$|\eta|\lesssim
1.0$}} along with the 
predictions from the GMSB model with a $\CONE$ mass of 175~GeV/$c^2$, normalized to 
202~$\invpb$.
The diphoton candidate sample data are in good agreement 
with the background predictions. There are no events above the 
\mett\ $>$~45~GeV threshold. The properties of the two candidates above 40 GeV appear consistent with the 
expected backgrounds.
} 
\label{Data Plot}
\end{centering}
\end{figure}

%
%
\mysect{}

In conclusion, we have searched 202~pb$^{-1}$ of inclusive diphoton events at 
CDF run II for anomalous production of missing transverse energy as evidence of 
new physics. We find good agreement with standard model expectations. We find no 
events above the {\it a priori} \mett\ threshold, and thus observe no new 
$ee\gamma\gamma\mett$ candidates. Using these results, we have set limits on the 
lightest chargino M$_{\cone}>167$ GeV/$c^2$ and \mbox{M$_{\none}>93$~GeV/$c^2$} 
at 95\% C.L. in a GMSB model. This limit is an improvement over previous CDF and 
D\O\ limits and is comparable to LEP II for similar models~\cite{OtherResults}.

%
%
\mysect{}

We thank the Fermilab staff and the technical staffs of the participating institutions for their vital contributions. This work was supported by the U.S. Department of Energy and National Science Foundation; the Italian Istituto Nazionale di Fisica Nucleare; the Ministry of Education, Culture, Sports, Science and Technology of Japan; the Natural Sciences and Engineering Research Council of Canada; the National Science Council of the Republic of China; the Swiss National Science Foundation; the A.P. Sloan Foundation; the Bundesministerium fuer Bildung und Forschung, Germany; the Korean Science and Engineering Foundation and the Korean Research Foundation; the Particle Physics and Astronomy Research Council and the Royal Society, UK; the Russian Foundation for Basic Research; the Comision Interministerial de Ciencia y Tecnologia, Spain; and in part by the European Community's Human Potential Programme under contract HPRN-CT-2002-00292, Probe for New Physics.


\begin{table}[hb]
\begin{tabular}{c|c|c|c|c||c}
\multicolumn{1}{c|}{\mett\ }   & \multicolumn{4}{c||}{Expected} & Observed \\ \cline{2-5}
Requirement &  QCD & $e\gamma$ & Non-Collision & Total &  \\ 
\hline\hline
25 GeV  & $4.01\pm 3.21 \pm 3.76$ & $1.40 \pm 0.52 \pm 0.45$ & $0.54 \pm 0.06 \pm 0.42  $ & $5.95\pm 3.25 \pm 3.81$ & 3 \\ \hline
35 GeV  & $0.30\pm 0.24 \pm 0.22$ & $0.84 \pm 0.32 \pm 0.27$ & $0.25 \pm 0.04 \pm 0.19  $ & $1.39\pm 0.40 \pm 0.40$ & 2 \\ \hline
45 GeV  & $0.01\pm 0.01 \pm 0.01$ & $0.14 \pm 0.06 \pm 0.05$ & $0.12 \pm 0.03 \pm 0.09  $ & $0.27\pm 0.07 \pm 0.10$          & 0 \\ \hline
55 GeV  & (negligible)            & $0.05 \pm 0.03 \pm 0.02$ & $0.07 \pm 0.02 \pm 0.05  $ & $0.12\pm 0.04 \pm 0.05$         & 0 \\
\end{tabular}
\caption{Numbers of events observed and events expected from background sources 
as a function of the \mett\ requirement. Here ``QCD'' includes the $\gamma\gamma$, 
$\gamma j$ and $jj$ processes.  The first uncertainty is statistical, the second is
systematic. }
\label{found}
\end{table}


\begin{thebibliography}{99}
\bibitem{defs}
We use a cylindrical coordinate system that defines $z$ as the longitudinal 
axis and along the proton beam axis, in which $\theta$ is the
polar angle, $\phi$ is the azimuthal angle and $\eta=-\ln \tan(\theta/2)$. In general, all quantities are defined 
from $z_{\rm vertex} = 0$, $E_T=E\sin\theta$ 
and $p_T = p \sin \theta$ where $E$ is the energy measured by the calorimeter and $p$ the 
momentum measured in the tracking system. $\vec{\met}=-\sum_i E_T^i \vec{n_i}$ 
where $\vec{n_i}$ is a unit vector that points from the interaction vertex to 
the $i$th calorimeter tower in the transverse plane.
$\met$ is the magnitude of $\vec{\met}$.
\bibitem{Fiducial} 
The fiducial region has $\sim$87\% coverage in the central
region.
\bibitem{iso} 
To reject hadronic backgrounds that fake prompt photons, candidates 
are required to be isolated in the calorimeter and tracking chamber. 
In the calorimeter the isolation is defined as the energy in 
a cone of 0.4 in $\eta-\phi$ space, minus the photon cluster energy, 
and corrected for energy loss into cracks as well as the number of reconstructed
$\ppbar$ interactions in the event.  We require isolation $<0.1\times\Etg$ for $\Etg<$~20~GeV, 
and $<2.0$~GeV$+0.02\times(\Etg -20$~GeV) for $\Etg>$~20~GeV. In the tracking chamber 
we require the scalar sum of the \ptt\ of all tracks in a cone of 0.4 
to be $<2.0$~GeV$+0.005\times\Etg$,  where all values of $\Etg$ are in GeV.
\bibitem{ETOUT} We require the time of arrival of the energy in all hadron calorimeter towers with at
least 0.5~GeV to be within 3$\sigma$ of the expected value.
\bibitem{LooserCuts} The identification and isolation requirements for the control sample are:
a)~isolation $<0.15\times\Etg$ for $\Etg<$~20~GeV, 
and $<3.0$~GeV$+0.02\times(\Etg -20$~GeV) for $\Etg>$~20~GeV;
b)~tracking isolation $<5$~GeV/$c$;
c)~Had/EM $<0.125$;
d)~at most one track, and no tracks with \mbox{\ptt$~>0.25~E_T/c$}.
\bibitem{MetMethod} 
The means of the $\Sigma \Et$ distributions (where $\Sigma \Et$ excludes the photon $\Et$'s)
are separated by approximately 6\%. This is most likely due to
different fractional contributions from $\gamma\gamma, \gamma j$
and $jj$ processes. The corrections are made by taking the expected shape of the
\mett\ distribution from the control sample as a function of $\Sigma
\Et$, and normalizing to the observed $\Sigma \Et$ distribution in the
signal sample.
\end{thebibliography}
\end{document}